**Insight into properties of sizable glass former from volumetric measurements**

Marzena Rams-Baron[1*], Alfred Błażytko[1], Riccardo Casalini[2], Marian Paluch[1]

[1]*August Chełkowski Institute of Physics, University of Silesia, 75 Pulku Piechoty 1, 41-500 Chorzow, Poland*

[2]*Chemistry Division, Naval Research Laboratory, 4555 Overlook Avenue SW, Washington, DC 20375, USA*

*corresponding author: marzena.rams-baron@us.edu.pl

**Abstract**

Sizable glass formers feature numerous unique properties and potential applications, but many questions regarding their glass transition dynamics have not been resolved yet. Here we analyzed structural relaxation times measured as a function of temperature and pressure in combination with the equation of state obtained from pressure-volume-temperature (PVT) measurements. Despite evidence from previous dielectric studies indicating a remarkable sensitivity of supercooled dynamics to compression, and contrary to intuition, our results demonstrated the temperate proof for the almost equivalent importance of thermal energy and free volume fluctuations in controlling reorientation dynamics of sizable molecules. The found scaling exponent $\gamma = 3.0$ and $E_v/E_p$ ratio of 0.6 were typical for glass-forming materials with relaxation dynamics determined by both effects with a minor advantage of thermal fluctuations involvement. It shows that the high values of key parameters characterizing the sensitivity of the glass transition dynamics to pressure changes, i.e. activation volume $\Delta V$ and $dT_g/dP$, are not a valid premise for a remarkable contribution of volume on glass transition dynamics.

The dynamics of molecular reorientation play an important role in many classes of materials, determining their physical and mechanical properties and underlying applications. Recently the behavior of glass-forming materials being a collection of π-conjugated heterocyclic units attracted our attention, in part because the reorientation dynamics of large, rigid, and anisotropic systems are novel in comparison to "traditional" glass-formers. [1], [2], [3], [4] Such materials occupy an underexplored region between more thoroughly researched low-molecular weight glass formers and larger but flexible polymers and are likely to display directional dynamic properties marked with inertial effects. Furthermore, their complex structure with many ring-like units connected in a way that ensures the optimal path for electron transport can be useful in various potential optoelectronic applications. Thus, understanding their reorientation dynamics is of both fundamental and applied interest.

Our previous ambient pressure dielectric relaxation research revealed that the novel properties observed for sizable and anisotropic glass formers are determined by the size and rigidity of their molecular cores (substantial inertial moments). [1] For instance, when we compared the temperature dependence of the structural relaxation times for a sizable system, with one of the widely studied low-molecular weight glass former i.e. propylene carbonate (PC), the differences in a high-temperature limit of the Vogel–Fulcher–Tammann (VFT) [5], [6], [7] fits were noticed (see Fig. 1a). This effect can be attributed to different inertial moments, I, as the pre-exponential factor in the VFT equation $\tau_0 \sim (2\pi I/k_B T)^{0.5}$ depends on I (T is temperature and $k_B$ is a Boltzmann constant). [8] For PC $I = 4.90 \cdot 10^{-45}$ kg·m² while for sizable molecule M-*para*-CF$_3$ $I = 3.19 \cdot 10^{-43}$ kg·m². It yields 8 times higher $\tau_0$ for a sizable system (see Fig.1a). Another consequence of their sizable structure is that their reorientation is not uniform in all directions and depends on the mass distribution relative to the reorientation axis. Due to the elongated shape, the inertial effects variously affect the time scale of short and long axes reorientations leading to complex dielectric response at high temperatures, in a less cooperative dynamic regime. [1], [3]

These new effects observed in dielectric research at ambient pressure led to another important question about the behavior of large, rigid, and anisotropic molecules under elevated pressure, namely, to what extent, and which parameters characterizing the dielectric response of sizable molecules to the applied pressure will reflect their large size. Jędrzejowska et al. reported a striking pressure sensitivity of glass transition dynamics for a representative of a family of sizable glass-forming molecules. [4] It was found that a change in $T_g$ of more than 40 K could be achieved by applying relatively low-pressure values such as 100 MPa. The resultant pressure coefficient of glass transition, $dT_g/dP$, was much higher than that of other conventional glass-forming systems, including polymers that are known to feature a pronounced sensitivity of $T_g$ to hydrostatic compression. [9] The second characteristic feature revealed by our high-pressure experiments were large values of activation volume which proves the extraordinary sensitivity of the sizable molecule's reorientation to the applied pressure. [3]

The above-mentioned effects prove that the dynamics of sizable glass formers are different from the dynamics of ordinary molecular and polymeric glass formers. Therefore, it is interesting to find out which thermodynamic quantity controls the unusual reorientation properties of sizable molecules in a supercooled liquid state. Many models have been proposed to describe the slowdown of reorientation dynamics on approaching the glassy state, including free volume models [10], [11], or energy landscape models [12], [13], [14], [15], [16] that selectively focus on temperature or volume as the main drivers of supercooled dynamics. On the other hand, experimental results showed that the increase in the characteristic relaxation time near $T_g$ for most glass-forming materials is due to both effects, i.e., densification enhancing the molecular crowding and loss of thermal energy leading to the entrapment of molecules in energy minima. [17], [18] Despite many decades of research advancing the understanding of glass transition, no generally accepted physical explanation has been provided for the mechanism underlying the mysterious slowdown of molecular dynamics on approaching $T_g$ which would satisfactorily take into account the role of temperature and volume with all empirical constraints. Consequently, further theoretical and experimental efforts are required to disentangle the role of both factors. For example, no group of materials has yet been identified that would allow for rigorous verification of existing free-volume theories. For this purpose, it is necessary to identify a class of materials for which packing effects play a significant role and volume is the main parameter controlling the dynamics. New insight can be provided by the class of sizable glass formers whose reorientation dynamics are unusual compared to liquids or polymers investigated so far. However, it must be determined first whether decreasing volume or thermal energy loss (or both) governs their glass-transition dynamics.

An important achievement bringing us closer to a complete understanding of the glass-transition dynamics was the observation that various dynamic quantities (such as structural relaxation time or viscosity) measured at different T-P conditions if plotted against $(TV^\gamma)^{-1}$ can be scaled to one master curve with only one adjustable parameter, the scaling exponent $\gamma$, being a material-specific constant. [19] This universal behavior was found for a wide range of different liquids with some exceptions (e.g. those with a significant hydrogen-bond network). [20], [21], [22], [23], Thus, a consistent theory of the glass transition should take this into account. It is interesting to see how the density scaling concept will work for sizable molecules with complex reorientation behavior. It is worth checking whether their large size and sensitivity of glass-transition dynamics to compression will affect the value of the $\gamma$ parameter, which depends on the relative contribution of temperature and volume to reorientation dynamics. In the case of materials whose dynamic properties are determined solely by free volume and density effects, a large value of the scaling exponent is expected. On the contrary, a small value of $\gamma$ can be anticipated when thermal fluctuations play an overwhelming role. [19] Along these same lines, it is interesting to determine whether the reorientation dynamics of sizable molecules around short and long molecular axes will exhibit a unique value of γ.

Our motivation behind the present study was to gain new insight into the atypical relaxation properties of sizable glass-formers by analyzing the structural relaxation times measured with dielectric spectroscopy as a function of temperature and pressure along with the equation of state established from pressure-volume-temperature (PVT) measurements. The combination of dielectric and PVT data allows us to disentangle the relative effects of T and V on the reorientation dynamics of a sizable system, which is of fundamental importance for understanding the role of their large molecular size and volume in reorientation dynamics control. The purpose of our study was 2-fold: (i) first, to verify which thermodynamic variable (volume, temperature, or both) has a dominant impact on molecular dynamics for a representative of sizable glass-formers; (ii) second, to perform a density scaling procedure to assess the scaling exponent γ and to compare the magnitude of γ for different aspects of the sizable molecule reorientation.

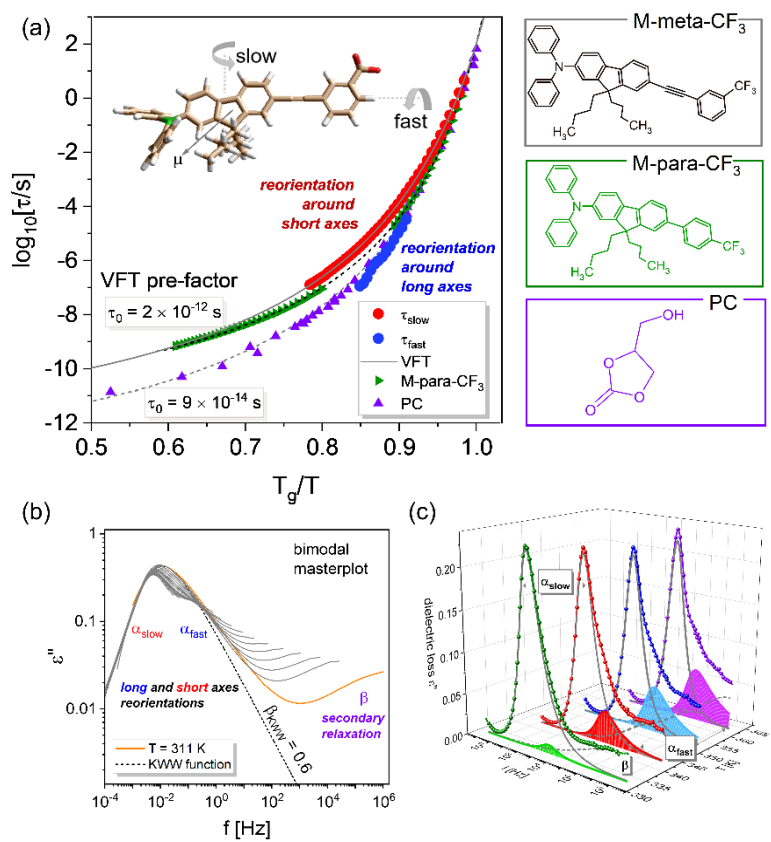

**Fig.1** (a) The comparison of the temperature dependence of relaxation times determined from dielectric spectroscopy for two representatives of sizable glass formers, i.e. M-para-CF$_3$ and M-meta-CF$_3$ (studied herein) and a low-molecular-weight glass-former – PC; right-site panels show their chemical structures. The differences in high-temperature limits of VFT fits are indicated. Data for M-para-CF3 and M-meta-CF3 were taken from ref. [1] and ref. [3]. Data for PC were taken from ref. [24] (b) Bimodal masterplot obtained by superimposing the dielectric loss spectra recorded for M-meta-CF$_3$ for temperatures ranges from 311 K to 379 K at ambient pressure. (c) Slow and fast contributions to dielectric loss spectra originated from short and long axes reorientations of the M-meta-CF$_3$ molecule.

The chemical structure of a sizable glass former selected for the present study referred to as M-meta-CF$_3$, is shown in Fig.1a. Its structure includes a sizable non-polar core (with dipole moment $\mu = 0.5$ D) being a collection of different ring-containing units, to which a phenyl ring substituted with –CF$_3$ group at meta position was attached. To improve the formation of a stable glassy phase, flexible alkyl chains were attached to the central unit of a sizable core. The orientation of the molecular (net) dipole moment of $\mu = 4.48$ D is shown in Fig. 1a (left panel). The parallel and perpendicular components of the dipole moment vector were $\mu = 3.92$ D and $\mu = 2.19$ D, respectively. The non-zero values of both components allowed us to probe dielectrically all aspects of the motion of the M-meta-CF$_3$ molecule. This is contrary to the case of M-para-CF$_3$, also taken into account in the analysis presented in Fig. 1a, where the dipole moment was directed parallel to the main molecular axis providing partial insight into whole molecule motions (only short axes reorientations were dielectrically active). The dielectric relaxation properties of M-meta-CF$_3$ at ambient and elevated pressure were investigated in detail elsewhere.[3] Its supercooled dynamics share common characteristics with representatives of sizable glass formers investigated to date. [1], [3], [4], [2] Far from T$_g$, in the high-temperature regime where the structural relaxation was less cooperative, the main dielectric loss peak in M-meta-CF$_3$ bifurcates revealing two modes resulting from reorientation around different molecular axes (see bimodal master plot in Fig.1b). [1], [3] Upon cooling towards the glass transition, which for M-meta-CF$_3$ was observed at T$_g$ = 308 K (from BDS, for $\tau = 100$ s), the ability to detect individual reorientation modes becomes limited as the cooperativity of supercooled dynamics increases. The contribution of individual relaxation modes to overall dielectric response for selected T in a less cooperative regime is shown in Fig. 1c. To make the evaluation clear, we collected dielectric loss spectra in a broader frequency range (up to GHz) in comparison to our previous [3] report using Agilent 4291B impedance analyzer connected with the Novocontrol GMBH system. The slow (low-$f$ component) and fast (higher-$f$ component, shaded with color in Fig.1c) contributions to the main relaxation peak were assigned to the motions dominated by reorientation around the short and long molecular axis, respectively. The former had a greater amplitude which is consistent with the larger magnitude of the parallel component of μ attempting this mode. The relaxation times $\tau_{slow}$ and $\tau_{fast}$ corresponding to short and long-axis reorientations exhibited non-Arrhenius T-dependence that obeys the VFT equation. [5], [6], [7] The prefactor $\tau_0$ in the VFT fit of the temperature dependence of $\tau_{slow}$ was found to be equal to $2 \times 10^{-12}$ s (see Fig.1a). This value was much longer than values typically observed for ordinary glass-forming liquids (where $\tau_0 \sim 10^{-14}$ s), which is due to large size and relevance of inertial effects. [1]

Volumetric measurements, providing access to the equation of state (EoS), gave us new insight into unusual relaxation properties revealed for sizable glass formers in our previous dielectric studies. Fig. 2a shows the temperature dependence of specific volume, V, measured for M-meta-CF$_3$ for various pressures (i.e. 10 - 50 MPa, Δp = 10 MPa). The relatively small pressure range included in the analysis resulted from the instability of M-meta-CF$_3$, which crystallized at higher pressures. From a step change in V(T) dependencies which evidenced a glass transition, the $T_g(P_g)$ line was estimated. As compared in Fig. 2b the $T_g(P_g)$ line determined from PVT (from the crossover of fitting functions to V(T) data) and BDS (from extrapolation of pressure VFT fits to τ = 100 s) were in reasonable agreement. The corresponding high value of the pressure coefficient of glass transition, i.e. $dT_g/dP$ = 390 K/GPa determined from Anderson-Anderson fit [25] to BDS data (as $k_1/k_3$ ratio), confirmed the supersensitivity of M-meta-CF$_3$ glass-transition dynamics to compression as reported in a previous paper. [3]

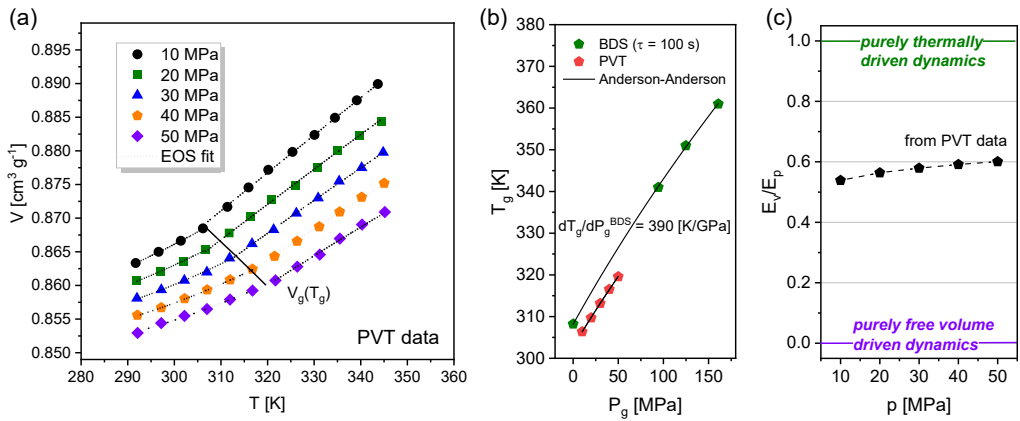

**Fig. 2** (a) Temperature variations of the specific volume V measured for M-meta-CF$_3$ at constant pressure for various pressures ranging from 10 MPa to 50 MPa. (b) The comparison of $T_g(P_g)$ lines determined from dielectric and PVT data. Note that the cooling rates for the volumetric and dielectric measurements were not the same. Solid lines correspond to the Anderson-Anderson [25] fitting function $T_g(p) = k_1(1 + p\, k_2/k_3)^{1/k_2}$ with $k_1$ = 308.2 ± 0.5 K, $k_2$ = 3.01 ± 0.6 , $k_3$ = 790.8 ± 37.9 MPa (c) $E_v/E_p$ calculated for M-meta-CF$_3$ from PVT data using eq.4.

To determine the scaling exponent γ, we parametrize the isobaric V(T) data collected above $T_g$ using the equation of state (EoS) [26]:

$$\left[\frac{v(T,p_0)}{v(T,p)}\right]^{\gamma^{EOS}} = 1 + \frac{\gamma^{EOS}(p-p_0)}{B_T(p_0)} \quad (1)$$

where $B_T(p_0) = b_1 exp[-b_2(T-T_0)]$ and $v(T,p_0) = A_0 + A_1(T-T_0) + A_2(T-T_0)^2$. The dotted lines depicted in Fig. 2a denotes best global fitting functions obtained for the following fitting parameters: $A_0$ = 0.875 cm$^3$/g, $A_1$ = 6.053 × 10$^{-4}$ cm$^3$/(g K), $A_2$ = 2.802 × 10$^{-9}$ cm$^3$/(g K$^2$), $b_1$ = 1817.18 MPa, $b_2$ = 8.25 × 10$^{-3}$ K$^{-1}$, and $\gamma^{EoS}$ = 14.97. The parameters $p_0$ = 0.1 MPa and $T_0$ = 308 K were fixed as constants. In turn, in the next step, we could express relaxation times measured dielectrically along isobar (0.1 MPa) and various isotherms (at 343 K, 353 K, 363 K) as a function of specific volume, V, using the EoS parameters. We took advantage of relaxation times for slow and fast reorientation modes, $α_{slow}$ nad $α_{fast}$, measured for M-meta-CF$_3$ at various T-P conditions elsewhere [3]. To assess scaling exponent γ, we used the method proposed by Casalini and Roland [27], providing the value of γ directly from fitting dielectric relaxation data presented in 3D space as a function of T and V with the Avramov model: [28]

$$log\tau(T,V) = log\tau_0 + \left(\frac{A}{TV^\gamma}\right)^D \quad (2)$$

where $log\tau_0$, A, D, and γ were fitting parameters. Fig.3a shows the relaxation times $log\tau_{slow}$(T, P) depicted in a T-V surface together with the outcome of the fitting procedure employing eq.2 (see Fig.3 caption for fitting parameters). From the best fit, the value of the scaling exponent was found to be γ = 3.0 for the $α_{slow}$ identified with short-axes reorientations. Knowing the value of γ, we could check the validity of the density scaling procedure for M-meta-CF$_3$. As demonstrated in Fig.3b the density scaling was satisfied for the $α_{slow}$ process. The relaxation times measured isothermally and isobarically at various T-P conditions in the supercooled liquid state superimposed well into a single line when plotted against $(TV^\gamma)^{-1}$ with γ=3.00 ± 0.03. Since we collected two data sets for the M-meta-CF$_3$, one for each reorientation mode, in Fig.3b two mastercurves are presented. The second one included relaxation times $\tau_{fast}$(T, P), resulting mostly from long-axis reorientations, available in a smaller T-V range in comparison to $α_{slow}$. Interestingly, the corresponding relaxation data could be satisfactorily scaled with the same value of the γ parameter, i.e. γ = 3.0 as for the $α_{slow}$. A slightly better scaling outcome could be obtained for γ = 3.15 ± 0.1 as indicated by free fit employing eq. 2. (see inset to Fig.3a). Regardless of the γ used in the analysis, obtained results indicated that both relaxation modes,

α$_{slow}$, and α$_{fast}$, in M-meta-CF$_3$ scale with an almost identical scaling exponent γ ≈ 3. This means that both types of mobility are determined by the same intermolecular potential.

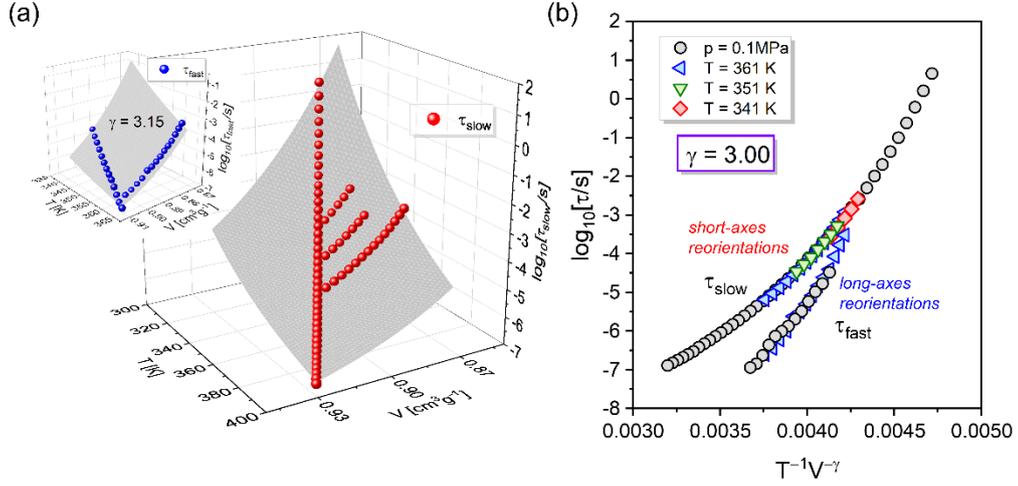

**Fig. 3** (a) Isothermal and isobaric relaxation times collected for α$_{slow}$ and α$_{fast}$ (inset) plotted as a function of temperature and volume. Grey areas represent surface fits according to eq.2. The corresponding fitting parameters are logτ$_0$ = -8.3 ± 0.1 s, A = 338 ± 2, D = 4.7 ± 0.1, and γ = 3.00 ± 0.03. For data presented in inset: logτ$_0$ = -8.9 ± 0.4 s, A = 297 ± 12, D = 6.7 ± 0.9, and γ = 3.15 ± 0.1. (b) Experimental verification of the density scaling rule for α$_{slow}$ and α$_{fast}$. with γ = 3.0.

What was surprising here, at least from the perspective of our hypothesis about the possible important role of free volume in relaxation dynamics control, was the low value of the scaling exponent γ. The γ value can be treated as an indicator of the importance of volume and temperature in controlling relaxation phenomena for a given material. In borderline cases, when γ = 0, the temperature is the major variable driving the glass-transition dynamics, while when γ→∞, density is getting more importance. [19] In all the known simple glass formers (i.e. without hydrogen bonds or an extensive covalent structure like polymers) the value of γ is found to be equal to or larger than 4 [29], thus it is quite unexpected that this class of molecules would have a γ significantly smaller than 4. The low scaling exponent found for M-meta-CF$_3$ excludes that volume is a principal factor controlling the unusual reorientation dynamics of sizable molecules. With relaxation properties clearly determined by density effects, a higher scaling exponent would be expected.

Another parameter that quantifies the relative contribution of T and V to glass-transition dynamics is the $E_v/E_p$ parameter, which could be defined as the ratio of the activation energy at constant volume $E_v = R[(\partial \ln\tau)/(\partial T^{-1})]_v$ to the activation enthalpy at constant pressure, $E_p = R[(\partial$ $\ln \tau)/(\partial\, T^{-1})]_p$. [17] The value of $E_v/E_p$ would be close to unity if the relaxation process is purely thermally activated and close to zero if dominated by density. Casalini and Roland showed an inverse correlation between $E_v/E_p$ and $\gamma$ as indicated by the following relationship [19]:

$$\left.\frac{E_v}{E_p}\right|_{T_g} = \frac{1}{1 + \alpha_p T_g \gamma} \qquad (3)$$

where $\alpha_p = V^{-1}\,[(\partial V)/(\partial T)]_p$ is the expansion coefficient at constant pressure. At $T_g = 308$ K and for $\gamma = 3.00 \pm 0.03$ the $E_v/E_p$ ratio calculated for $\alpha_{slow}$ in M-meta-CF$_3$ using eq.3 was equal to 0.61. It means that the contribution of temperature to the reorientation dynamics of an investigated sizable molecule could be even greater than that of volume, but both factors are important. Alternatively, the $E_v/E_p$ ratio can be calculated using the ratio of the isobaric $\alpha_p$ and isochronic $\alpha_\tau = V^{-1}\,[(\partial V)/(\partial T)]_\tau$ thermal expansion coefficients as [25]:

$$\left.\frac{E_v}{E_p}\right|_{T_g} = \frac{1}{1 + \alpha_p/|\alpha_\tau|} \qquad (4)$$

We used eq.4 to determine the $E_v/E_p$ ratio for M-meta-CF$_3$ from PVT data. For pressures between 10 MPa and 50 MPa, we found out that the $E_v/E_p$ varies from 0.54 to 0.60, see Fig. 2c, indicating that both T and V almost equally impact the relaxation dynamics in M-meta-CF$_3$. The obtained results meet the anticorrelation reported for the $E_v/E_p$ ratio and $\gamma$ parameter for a diverse group of distinct glass-forming molecules. [19], [21] Still, the lowest $E_v/E_p$ value observed to date was found for BMPC and BMMPC ($E_v/E_p \approx 0.4$ and $\gamma > 7$) pointing out that volume is a key factor controlling their glass transition dynamics. [17]

Our results are interesting in the context of the discussion about molecular factors determining the relative contribution of thermal and density fluctuations in controlling the dynamics, and parameters that for a given system might provide useful insights into these fundamental properties, accurately indicating the prominent impact of density on relaxation times. The extraordinary sensitivity of M-meta-CF$_3$ glass transition dynamics to compression, reflected in the high value of the $dT_g/dP$ parameter, cannot be taken as a guideline for this purpose. This claim can be supported by invoking the previously mentioned example of low molecular-weight glass former PC with a $dT_g/dP = 0.09$ K/MPa [30] which is more than four times lower than $dT_g/dP$ value found for M-meta-CF$_3$. The small molecule, PC, when compared to the studied sizable system has a very similar $E_v/E_p = 0.64$ [31], and an even slightly larger $\gamma$ ranging from 3.7 to 4.3. [31], [32], [33] Therefore, the hypersensitivity of the glass-transition dynamics to compression reported for sizable systems and expressed by a high $dT_g/dP$ value should not be identified as a premise for a remarkable contribution of volume on relaxation times dynamics, but rather as an effect of large size and massiveness of their chemical structures.

In summary, we have demonstrated here for the first time, the scaling properties of the dynamics (i.e. γ scaling) for a representative of sizable glass-forming molecules. Contrary to what we originally expected, the scaling exponent γ = 3.00 ± 0.03 and the $E_v/E_p \approx 0.6$ determined for M-meta-$CF_3$ suggested that the unusual and complex reorientation dynamics of a sizable molecule depends on both thermodynamic variables but surprisingly, temperature plays a more dominant role than volume. The presented evidence could be strengthened in future research on a representative of sizable glass formers with lower crystallization propensity, allowing for the analysis of volumetric and dielectric data over a wider range of pressures, temperatures, and volumes without the risk of sample crystallization. The results presented here expand on our previous research on the specific features of sizable glass formers bearing unusual dynamic properties determined by the massiveness, stiffness, and anisotropy of their chemical structure. Interestingly, the uniqueness of sizable molecules reflected in large pre-exponential factors in VFT function (*i*), high values of $dT_g/dP$ (*ii*), and large activation volumes (*iii*) was not accounted for by the density scaling outcome which shows quite universal features. The complex and remarkable pressure-sensitive reorientation dynamics in M-meta-$CF_3$ scales with the same and low value of γ parameter for both reorientation modes as is typically observed for materials with relaxation properties coupled to both temperature and density effects. Our data support the observations that for most glass-forming materials (with strongly associating liquids as an exception) the dielectric relaxation properties do not depend on a single thermodynamic quantity. Consequently, both T and V play a relevant role in governing structural relaxation times in the vicinity of $T_g$. The same is true for sizable glass formers.

**Acknowledgments**

This research was funded in a whole by the National Science Centre, Poland [Opus 21, 2021/41/B/ST5/00992]. For Open Access, the author has applied a CC-BY public copyright license to any Author Accepted Manuscript (AAM) version arising from this submission. The study was implemented as part of strategy of the University of Silesia—Inicjatywa Doskonałości (POB 1—Priorytetowy Obszar Badawczy 1: Harmonijny rozwój człowieka—troska o ochronę zdrowia i jakość życia). Author R. C. acknowledges the support of the Office of Naval Research (N0001423WX02080).